\documentclass[structabstract]{aa}  
\usepackage{graphicx}
\usepackage{txfonts}
\usepackage{natbib}
\bibpunct{(}{)}{;}{a}{}{,} 

\begin{document}

   \title{Data Reduction Techniques for High Contrast Imaging Polarimetry}

   \subtitle{Applications to ExPo}

  \author{H. Canovas, M. Rodenhuis, S. V. Jeffers, M. Min, C. U. Keller
          \inst{1}}

   \institute{Sterrekundig Instituut, Universiteit Utrecht, PO Box 80000, 3508 TA Utrecht, The Netherlands \\
              \email{H.CanovasCabrera@uu.nl}
             }

\date{Received 18 March 2011, Accepted 10 May 2011}

    \abstract
   {Imaging polarimetry is a powerful tool for detecting and characterizing exoplanets and circumstellar environments. 
   Polarimetry allows a separation of the light coming from an unpolarized source such as a star and the polarized source such 
   as a planet or a protoplanetary disk. Future facilities like SPHERE at the VLT or EPICS at the E-ELT will incorporate imaging polarimetry 
   to detect exoplanets. The Extreme Polarimeter (ExPo) is a dual-beam imaging polarimeter that currently can reach contrast ratios of $10^{5}$, 
   enough to characterize circumstellar environments.}
   {We present the data reduction steps for a dual-beam imaging polarimeter that can reach contrast ratios of $10^{5}.$}
   {The data obtained with ExPo at the William Herschel Telescope (WHT) are analyzed. Instrumental artifacts and noise sources are discussed 
   for an unpolarized star and for a protoplanetary disk (AB Aurigae).}
   {The combination of fast modulation and dual-beam techniques allow us to minimize instrumental artifacts. A proper data processing and alignment of
   the images is fundamental when dealing with large contrasts. Imaging polarimetry proves to be a powerful method to resolve 
   circumstellar environments even without a coronagraph mask or an Adaptive Optics system.}
 {}

\titlerunning{Data Reduction Techniques for High Contrast Imaging Polarimetry}
\authorrunning{Canovas et al.}
\maketitle

\section{Introduction}

The direct detection and characterization of exoplanets is one of the main goals of the next generation of instruments and telescopes, such as the E-ELT \citep{Hook}.
Despite the increasing amount of detected exoplanets (e.g., \textit{http://www.exoplanet.eu/}), only a few of them have been directly imaged 
\citep{Marois28112008, Kalas2008,Lagrange02072010,Todorov05052010}.

Polarization is a powerful tool for detecting and characterizing circumstellar environments, such as protoplanetary disks, 
debris disks and exoplanets \citep{kuhn_2001,Stam_2005,Stam_2006}. While the light coming from the star is largely unpolarized, the scattered light from exoplanets and 
circumstellar environments is highly (linearly) polarized. This makes it easy to separate the planet signal from the unpolarized starlight, and reach the required, large contrast
ratios.

Future exoplanet instruments such as SPHERE \citep{sphere_2008}, GPI \citep{gpi_2007} or EPICS  \citep{epics_2010} will include imaging polarimetry, spectropolarimetry, 
or both, to characterize exoplanets. To reach the high contrast ratios that are necessary to detect an exoplanet, these instruments will combine coronagraphs,
Adaptive Optics and polarimetry. The combination of the firsts two elements can reduce the stellar light by a factor of $10^{6}$ \citep{coronograph_sphere}, while
polarimetry alone can reach contrast ratios of $10^{5}$ \citep{Rodens_2011}. The identification of instrumental artifacts and spurious signals that appear at these extremely 
high ratios is fundamental to understand the data produced by the new generation of exoplanet imagers. 

ExPo is a dual-beam imaging polarimeter that combines speckle imaging with dual-beam techniques, currently reaching contrast
ratios of $10^{5}$  without a coronagraph or adaptive optics system \citep{Rodens_2008,Rodens_2011}. The sensitivity reached by this instrument 
allows us to image circumstellar environments such as protoplanetary disks or dust shells with unprecedented accuracy 
\citep[e.g.][]{Canovas_2011_b,Canovas_2011_a}

In this paper we present the data reduction approach for high dynamic range polarimetry. The whole data-reduction process is shown in
Fig.~\ref{fig:Pipeline Blocks}. Results show that the dual-beam technique 
effectively minimizes systematic errors, leading to sensitive polarization measurements of different circumstellar environments.

In the following sections, the instrument is briefly described,  followed by the analysis of the instrumental artifacts.  The image reduction process and a discussion of 
first results are presented at the end. Preliminary results of the well-known protoplanetary disk AB Aurigae and the diskless star HD12815 are shown.
\section{Instrument Description}
ExPo is a dual-beam imaging polarimeter that currently reaches contrast ratios of $10^{5}$. It is designed to characterize
circumstellar environments, such as  the disks around Herbig Ae/Be and T Tauri stars. ExPo is currently a visitor instrument at the William Herschel 
Telescope (WHT), where it has been used for four successful observing campaigns.

ExPo has a wavelength range from $400-900$ nm, and its field of view is $20"\times20"$. It works with exposure times of 0.028 seconds to minimize the effects of
atmospheric seeing. Systematic errors are minimized by modulating two opposite polarization states. The combination of dual-beam, short exposures
times and polarization modulation allows us to reach contrast ratios of $10^{5}$ without the aid of Adaptive Optics (AO) or coronagraphs. 

As a dual-beam instrument, the incoming beam of light is divided into two orthogonally linearly polarized beams. As a result of this, two simultaneous 
images with opposite polarization states can be recorded on the CCD at the same time.

\begin{figure*}
  \centering
\resizebox{\hsize}{!}{\includegraphics[trim = 0 400 0 10]{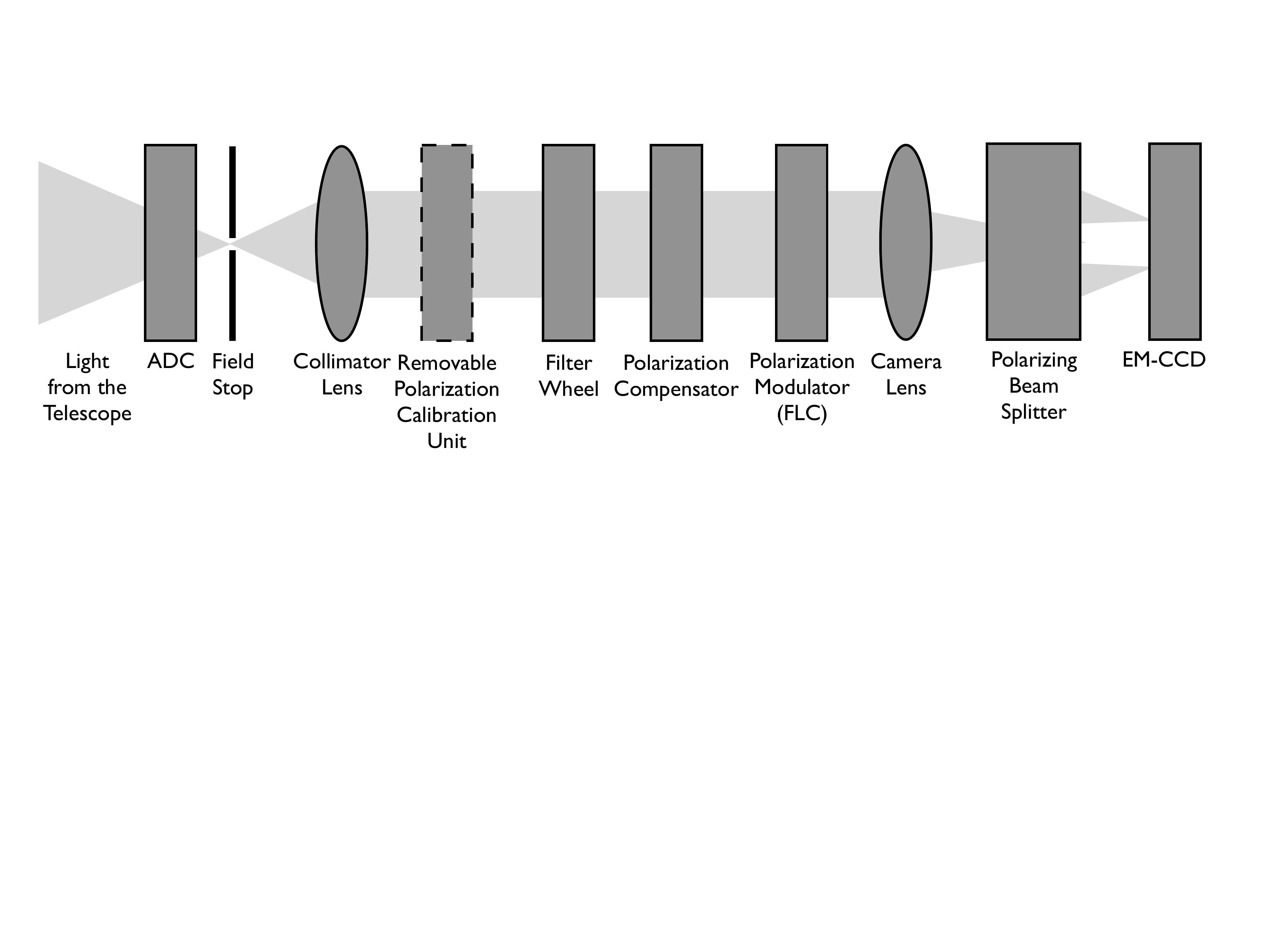}}
\caption{Schematic layout of the instrument in the Nasmyth focus of a telescope.}
\label{fig:ExPo_Layout}
\end{figure*}

The schematic layout of the instrument is shown in Fig.~\ref{fig:ExPo_Layout}. The light beam enters the instrument from the left. The dispersion introduced by the 
atmosphere \citep[e.g.][]{adc} is reduced by an Atmospheric Dispersion Corrector (ADC). The field stop prevents beam overlap on the detector. 
The filter wheel contains broadband and narrowband filters as well as an opaque filter to obtain dark measurements. The polarization compensator minimizes the 
instrumental polarization introduced by the telescope. The light beam is currently modulated by a single Ferroelectric Liquid Crystal (FLC), which switches 
between two states (A and B) with orthogonal polarization states. Future versions of this instrument may use an achromatic 3-FLCs Pancharatnam modulator 
\citep{Pancharatnam,Keller_2002}, but this will not influence the data reduction approach described in here.
The beamsplitter divides the incoming light into two beams (left and right) with perpendicular polarization states. These two beams are finally imaged onto 
the EM-CCD camera,  which records two simultaneous images with orthogonal polarization states. During the observation, the whole instrument remains static, 
with only the FLC switching from one state to the other.

\subsection{ExPo Images}
With the FLC in two different states, two different frames ($A$ and $B$) are recorded. A total of four images (i.e. two images per frame, corresponding
to the \textit{left} and \textit{right} beams) are produced for each FLC cycle: $A_{L},A_{R},B_{L},B_{R}$, where the subscripts 
L and R stand for left and right, respectively. Due to the FLC modulation, $A_{L}$ and $B_{R}$ contain the same polarization information, as happens 
to $A_{R}$ and $B_{L}$. To account for all the components of these images, we use the following notation: 

\begin{figure*}
  \center
   \includegraphics[width = 1.\linewidth,trim = 0 200 0 0 ]{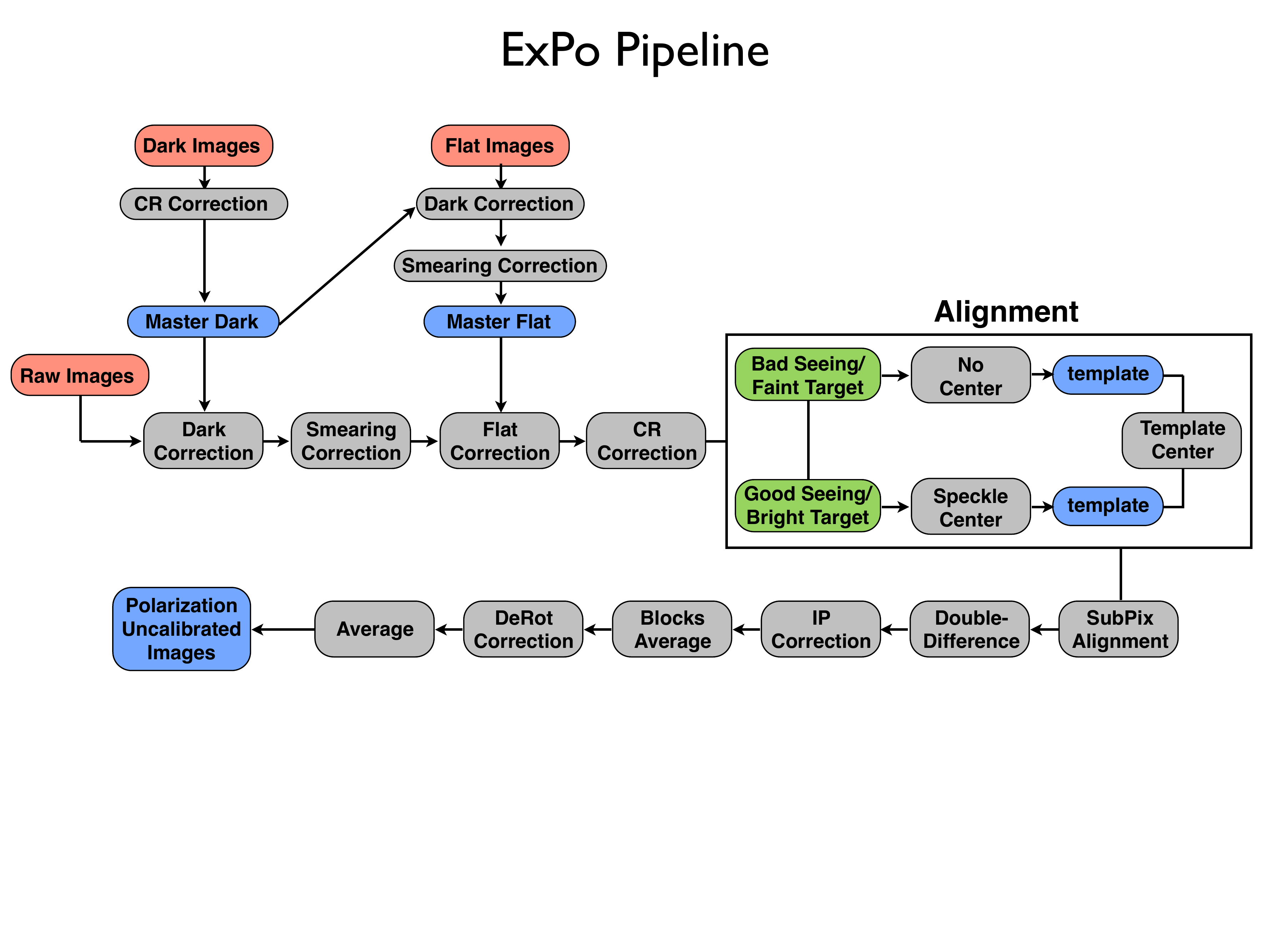}
   \caption{Pipeline Block Diagram.}
  \label{fig:Pipeline Blocks}
\end{figure*}

  \[
      \begin{array}{lp{0.8\linewidth}}
\mathrm{T_{A,B}}						& FLC transmission						\\
\mathrm{T_{L,R}}						& Beamsplitter transmission				\\
\textit{Flat}_\mathrm{{AL,AR,BL,BR}}	& Flat Fields								\\
I,P									& Intensity, Polarization images			\\
\textit{PSF}							& Point Spread Function					\\     
N									& Noise
      \end{array}
   \]
\noindent
where $\mathrm{T_{A,B}}$ accounts for the FLC transmission in its two possible states, and $\mathrm{T_{L,R}}$ accounts for any transmission difference
between the two beams. ${Flat}_\mathrm{AL,AR,BL,BR}$ accounts for the different flat fields on the left and right sides of the CCD for the A and B images.
The noise term includes the read-out and
the photon noise. The polarization image includes any polarized light that enters the instrument, including sky polarization and polarization introduced by the telescope. The PSF 
can be expressed as the convolution of the ``optics'' and ($\mathrm{PSF_{Opt}}$) and ``seeing" ($S$) terms of the PSF  \citep[e.g.][]{Adaptive}:
\begin{equation}
\mathrm{PSF} = \mathrm{PSF_\mathrm{{Opt}}} *S,
\end{equation}
where $\mathrm{PSF_{Opt}}$ accounts for the static optical elements of the instrument and telescope, and $S$ accounts for the dynamic
atmosphere. Any small imperfection in the beamsplitter, or difference in the optical path between the two beams will translate into different $\mathrm{PSF_{Opt}}$
for the left and right images: $\mathrm{PSF_\mathrm{{L}}}$ and $\mathrm{PSF_\mathrm{{R}}}$. On the other hand, the atmosphere will change with time, 
so $S$ will also change for the A and B frames: $S_\mathrm{A}$ and $S_\mathrm{B}$. Taking all of this into account, the four images recorded by the CCD in one FLC 
cycle can be described by:
\begin{eqnarray}
A_\mathrm{L}	&=& \mathrm{T_{A}} \cdot \mathrm{T_{L}} \cdot Flat_\mathrm{AL} \cdot 0.5 \cdot  \bigl((I +P)*\mathrm{PSF_{L}}*S_\mathrm{A} \bigr) + N_\mathrm{AL}  \nonumber \\
A_\mathrm{R}	&=& \mathrm{T_{A}} \cdot \mathrm{T_{R}} \cdot Flat_\mathrm{AR} \cdot 0.5 \cdot \bigl((I - P)*\mathrm{PSF_{R}}*S_\mathrm{A} \bigr) + N_\mathrm{AR} \nonumber \\
B_\mathrm{L}	&=& \mathrm{T_{B}} \cdot \mathrm{T_{L}} \cdot Flat_\mathrm{BL} \cdot 0.5 \cdot  \bigl((I - P)*\mathrm{PSF_{L}}*S_\mathrm{B} \bigr) +  N_\mathrm{BL} \nonumber \\
B_\mathrm{R}	&=& \mathrm{T_{B}} \cdot \mathrm{T_{R}} \cdot Flat_\mathrm{BR} \cdot 0.5 \cdot \bigl((I +P)*\mathrm{PSF_{R}}*S_\mathrm{B} \bigr) + N_\mathrm{BR},
\label{eq:imequation}
\end{eqnarray}
where $A_{L}$ stands for the left image taken when the FLC is in its A state, and so on.  $N_{AL}$ refers to the noise associated with this image.
Symbols ``$\cdot$" and ``*" stand for multiplication and convolution, respectively. The factor $0.5$ comes from the beamsplitter: the total incoming beam
is divided into two. A "pure" intensity image can then be generated by adding the left and right beams.  

Once a data set is recorded, the FLC is rotated by $22.5^{\circ}$, and a new data set is acquired. Since ExPo does not have a derotator, the field rotation 
will not only rotate the image, but also the polarization plane of the incoming light. To avoid this, the images are calibrated according to the reference 
system of the observed target, and then transformed to the reference system of the instrument. The Stokes parameters Q and U are obtained after 
combining and calibrating \citep{Rodens_2011} four datasets with the FLC rotated by $0^{\circ}$, $22.5^{\circ}$, $45^{\circ}$ and $67.5^{\circ}$ 
(e.g. \citet{Patat_2006}). Two FLC positions ($0^{\circ}$ and $22.5^{\circ}$ or $45^{\circ}$ and $67.5^{\circ}$) are necessary to produce calibrated images, 
so two redundant data sets are obtained after calibrating. Random errors (e.g. uncorrected cosmic rays, FLC ghosts) are removed by comparing these two data sets.

\section{Image Preparation}

In this section we describe the initial reduction steps that must be applied to the raw data. Once the images are corrected for dark, flat and smearing effects, they
are ready to be aligned and combined as explained in \S4 and \S5.

\subsection{Dark Reduction}
A dark image is an image taken with a given exposure time and with the CCD shutter closed. However, this is not possible due to the short exposure times required
by ExPo, as will be explained in the next subsection, and the CCD shutter must remain open during the whole observation.  
An opaque filter is placed in the filter wheel to measure dark frames.

Every dark frame is checked to detect and remove Cosmic Rays (CR). CR are identified as pixels showing values greater than a threshold value. This 
threshold is defined according to the characteristics of the dark observation, i.e. dark frames obtained with different CCD gain values will require different threshold 
values to detect the CR events. This process produces good results when reducing dark frames. Another approach is used when reducing the final science images,
as is explained in \S3.4. 
Finally, a very low-noise master dark is generated by combining at least 8000 dark frames (equivalent to a total integration time of roughly 4 minutes). 
\subsection{Smearing and Second-order Corrections}
Due to the short exposure times used by ExPo, the CCD runs in $frame$-$transfer$ mode, which means that the detector is reading and recording frames at the 
same time. To achieve this, the CCD chip is split in half: one half for exposing and another half for reading. Once an image is recorded on the ``exposing" half, 
is shifted to the read-out area. The shutter remains open during the whole process, and this, unfortunately, introduces spurious effects that must be removed 
in the initial reduction of the raw images.

Each time a frame is transferred to the read-out area of the CCD, there is some leftover charge. As a result of this, an extra
background is added to the next recorded frame. Moreover, in case the observed target is too bright, a streak in the direction of the charge shifting will appear when
exposing for a long time, or averaging many short-exposure images. 

\begin{figure}
  \centering
     \resizebox{\hsize}{!}{\includegraphics[trim = 30 500 50 60 ]{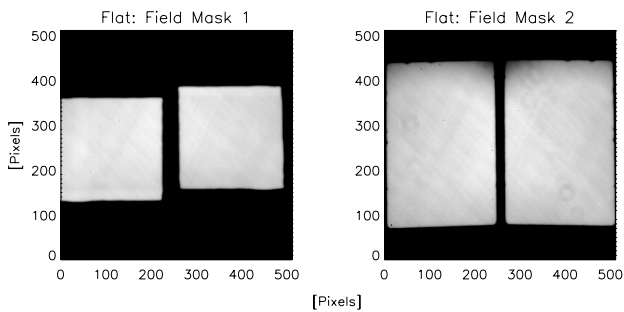}}

   \caption{ Normalized Flat Fields with two different field masks.}
  \label{fig:flats}
\end{figure}

\begin{figure}
  \centering
   \resizebox{\hsize}{!}{\includegraphics[trim = 30 500 50 60 ]{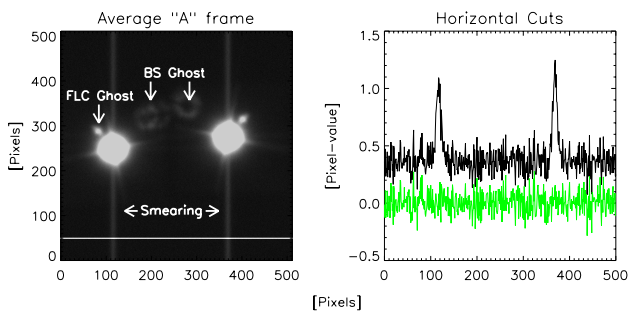}}
   
\caption{Left: Raw image, plotted at very high contrast. Two spurious ghosts can be seen above the star (discussed later in \S6.4) The horizontal white line represents the 
plot that is shown on the right. Right: The black line shows the horizontal cut before the correction, the green line shows the smearing correction effect described in \S3.2.}
  \label{fig:no_smearing}
\end{figure}

Due to the field mask, some areas of the CCD are not illuminated (Fig.~\ref{fig:flats}). Any signal measured in these non-illuminated 
regions must come either from the extra background mentioned above, or from the read-out noise. The information contained in these ``dark" pixels is used to remove these
two effects. An example of this is shown in Fig.~\ref{fig:no_smearing}: the spurious signal is first measured outside of the field mask,
and then subtracted from the raw image.
\subsection{Flat Reduction}

A set of dome flats is taken at the beginning and the end of each observing night. A minimum number of 100 frames are recorded for each 
of the four FLC positions ($0^{\circ}, 22.5^{\circ}, 45^{\circ}, 67.5^{\circ}$). Every frame is then corrected for dark and smearing effects. As with 
the images, it is convenient to distinguish between $Flat_\mathrm{AL}$, $Flat_\mathrm{AR}$, $Flat_\mathrm{BL}$ and $Flat_\mathrm{BR}$. Any polarization introduced 
by the dome lights, beamsplitter and FLC transmission differences disappears when normalizing each of these flats independently. All of the flat field 
images are combined to  generate four master flats.

The analysis of the flat fields shows that the CCD has no dead pixels. 
\subsection{Cosmic Ray Correction}

Before the images are combined and analyzed, they are corrected for cosmic rays (CR). A standard ExPo data set comprises 20000 images 
(around 10 minutes exposure time) per FLC position. The CR contribution cannot be neglected since we are dealing
with contrast ratios of $\approx 10^{5}$. Due to the nature of the ExPo images, with very short exposure times and a FOV of $20"\times20"$, 
it is not trivial to discriminate between a real speckle and a CR. Standard CR rejection methods are based on comparing an average image with 
the individual ones, and then apply a sigma-rejection algorithm. However, the speckle pattern disappears in the average; the average and 
individual images are indeed quite different. A different approach, based on the advantage of having two simultaneous images in the same frame, is used in 
this analysis. A box of $5\times5$ pixels is extracted around the brightest pixel of the left and right images.  The standard deviations ($\sigma$) of these two boxes are 
then compared. Experimental results show that, if the brightest pixel corresponds to a CR and not to a speckle, its associated 
$\sigma$ will be at least 3 times greater than the $\sigma$ of a true speckle. CR's are detected and corrected according to this threshold: if one of the boxes contains
a CR, all its pixels are set to the mean value of the whole image.

\begin{figure}
  \centering
\resizebox{\hsize}{!}{\includegraphics[trim = 90 430 75 145]{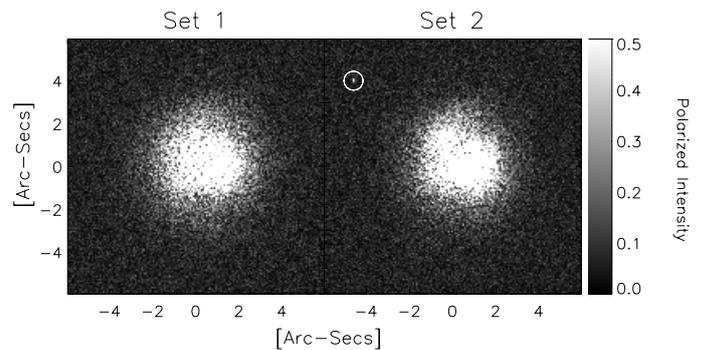}}
   \caption{Polarized images of AB Aurigae (seeing was $\sim 1.4"$). Set1 shows a calibrated image generated with the FLC placed at $0^{\circ}$ and $22.5^{\circ}$.
   Set 2 shows the polarized image generated after placing the FLC at $45^{\circ}$ and $67.5^{\circ}$. The white circle shows the position of a cosmic ray that does
   not appear in the first image.}
  \label{fig:cr}
\end{figure}

In case both the left and right images are affected by CR's, this method 
will still work because the probability that the cosmic rays have the same standard deviation is very low.
Fainter CR can be divided into two categories: the ones that fall inside the speckle pattern and the ones that fall outside of it. The former are, in practice, impossible 
to discriminate from true speckles. The latter are removed from the final images at the end of the analysis. Figure~\ref{fig:cr} shows an example of this process:
one uncorrected CR appears in one set but not on the other, so it can be easily identified and removed from the final image. 
\section{Image Alignment}
  
Once these corrections are applied, the images are aligned before they are combined. During the image acquisition, the telescope guides on the observed target (i.e.
there is not any \textit{reference star} to be tracked). As a result of this, the accuracy of the guiding decreases, specially in the fainter ($m_\mathrm{v} \geq 9)$ targets.
Because of the small field of view of ExPo, any small telescope misalignment can shift the image center by  several pixels. Additionally, due to the short exposures used, 
the PSF is a speckle pattern. All of these contributions make image alignment one of the main sources of error in the processing of ExPo data. To reduce misalignment 
effects, two centering algorithms were tested: brightest speckle and template cross-correlation.

The brightest speckle method (or shift-and-add method) is used in lucky imaging \citep[e.g.][]{luckyim}. The images are aligned according to the 
position of the brightest speckle in every image.

The template cross-correlation method first requires a starting template or reference image. Once the template is provided, each single 
image is cross-correlated with it. The image center is now defined as the position of the maximum value of this operation. Different templates such 
as a simulated gaussian-shaped Point Spread Function (PSF) or an average speckle-centered real PSF were used in these calculations. 

Once the images are centered, and before they are combined, they are again aligned to minimize any other source of misalignment, such as sub-pixel or flat-field
gradients. These effects are corrected with a custom version of the drizzle code of \citet{Drizzle}. In this code, each pixel is expanded into several pixels,
making it possible to shift and align images with sub-pixel accuracy.  

The analysis shows that the best results are obtained when combining these three methods. First, a template is generated by centering each left and right 
image according to the brightest speckle. An average non-gaussian shaped PSF is generated after
combining all these images. Then, this PSF is used as the input template for the second method: each individual image is now cross-correlated with this template.
Finally, each 'right' image is aligned with respect to the 'left' one, by using the drizzle code. Each pixel is subdivided into 9 sub-pixels, which results in an alignment
accuracy of a third of a pixel. Any shift introduced by the FLC is also corrected by this centering process.

\begin{figure}
  \centering
	\includegraphics[width=\columnwidth,trim = 80 330 50 45]{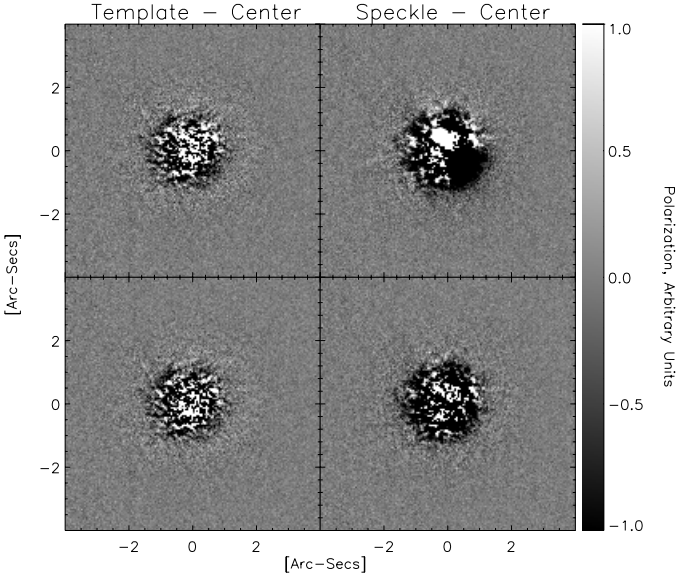}
   \caption{Polarized (uncalibrated) results for a diskless (unpolarized) star: HD122815. Four different alignment approaches are shown. Left column: images are
   aligned with a template. Right column: images re aligned according to the brightest speckle (shift-and-add). Top row: No sub-pixel 
   alignment. Bottom row: sub-pixel alignment. Color bar: amount of (uncalibrated) polarization, P.}
  \label{fig:center}
\end{figure}

To check how misalignment affects the data reduction, several unpolarized  (diskless) stars were analyzed. In these cases, no polarization pattern is expected to be
measured. The accuracy of the alignment can be quantified in terms of the standard deviation $\sigma$ of this pattern. 
Figure~\ref{fig:center} shows the polarization pattern
measured for the unpolarized star HD122815. The top row shows the difference between the template-center and the speckle-center. The bottom row shows the same result,
but now including sub-pixel alignment with a third of a pixel accuracy. The combination of template center plus sub-pixel alignment shows almost no structure at all
(Fig.~\ref{fig:center}, lower-left corner). 

From Fig.~\ref{fig:center} it is clear that most of the misalignment disappears when aligning the images according to the cross-correlation method. No significant 
improvement is detectable when aligning with sub-pixel accuracy due to the lack of structure in a diskless star.
\section{Image Combination}

\begin{figure}
  \centering
\resizebox{\hsize}{!}{\includegraphics[trim = 90 430 75 145]{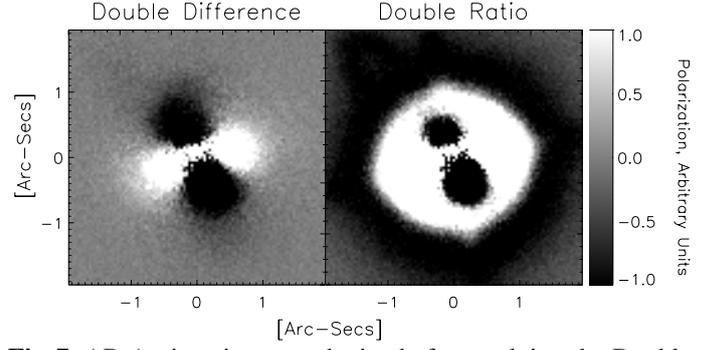}}
   \caption{AB Aurigae images, obtained after applying the Double Difference (left) and Double Ratio (right). Color Bar: amount of (uncalibrated) polarization, P.}
  \label{fig:doubleratio}
\end{figure}

As described by \citet{Bagnulo_2010} there are two methods  of combining images  produced with a dual-beam polarimeter to produce
polarization images: one based on the image \textit{ratios} and the other based on the image \textit{differences}. For ExPo, these two approaches 
can be expressed in terms of Eq.~\ref{eq:imequation} as
\begin{equation}
P = 0.25 \cdot \left(\frac{A_\mathrm{L}}{B_\mathrm{L}} \frac{B_\mathrm{R}}{A_\mathrm{R}} -1\right) \cdot 0.5 \cdot (A_\mathrm{L} + A_\mathrm{R} + B_\mathrm{L} + B_\mathrm{R})
\label{eq:dualrat}
\end{equation}
for the \textit{double-ratio} method, and as
\begin{equation}
P = 0.5 \left(P_\mathrm{A} - P_\mathrm{B}\right) = 0.5 \left(\left(A_\mathrm{L} - A_\mathrm{R}\right) - \left(B_\mathrm{L} - B_\mathrm{R}\right)\right)
\label{eq:doublediff}
\end{equation}
for the \textit{double-difference} method. Both methods were investigated. The results prove that better results are obtained when applying 
the \textit{double-difference} method instead of the \textit{double-ratio} explained in detail in \citet{keller_1996} and \citet{Donati1997}.
This result is shown in Fig.~\ref{fig:doubleratio}, where the same data set was reduced applying the \textit{double-differences} method (left) and 
the \textit{double-ratio} method (right). This can be explained by the difference of the speckle pattern for each image. Even though $A_\mathrm{L}$ and $A_\mathrm{R}$
are recorded simultaneously, they show small differences in their speckle patterns due to instrumental effects. Due to the huge dynamic range present in our
images, the slightest difference between two images will produce a big effect when calculating a ratio. However, the difference method is less sensitive
to this. Therefore, the latter method was chosen for ExPo. 
At this point the dual-beam + beam-exchange technique shows its advantages. The combination of the four images generated in each 
FLC cycle minimizes the systematic errors. Following the notation introduced in \S 2, all terms that appear when developing 
the equations are listed in Table~\ref{tab:DD Terms}.
\begin{table}
\label{tab:dd terms}
\centering
\begin{tabular}{ c  c  }
Notation						&	Expression									\\
\hline\hline
$\Delta A$					&	$A_\mathrm{L} - A_\mathrm{R}$				\\
$\Delta B$					&	$B_\mathrm{L} - B_\mathrm{R}$				\\
$\Delta L$					&	$A_\mathrm{L} - B_\mathrm{L}$				\\
$\Delta R$					&	$A_\mathrm{R} - B_\mathrm{R}$				\\
$\delta_\mathrm{I}$			&	$\mathrm{T_{L}\cdot PSF_{L} -T_{R}\cdot PSF_{R}}$ \\
$\delta_\mathrm{P}$			&	$\mathrm{T_{L}\cdot PSF_{L} +T_{R}\cdot PSF_{R}}$ \\
$\delta^{'}_\mathrm{I}$		&	$\mathrm{T_{A}}\cdot S_\mathrm{A} - \mathrm{T_{B}}\cdot S_\mathrm{B}$ \\
$\delta^{'}_\mathrm{P}$		&	$\mathrm{T_{A}}\cdot S_\mathrm{A} + \mathrm{T_{B}}\cdot S_\mathrm{B}$ \\
$\Delta N_\mathrm{A}$		&	$N_\mathrm{AL} - N_\mathrm{AR}$	\\ 
$\Delta N_\mathrm{B}$		&	$N_\mathrm{BL} - N_\mathrm{BR}$	\\
\hline
\end{tabular}
\caption{Notation used during the development of the image analysis.}
\label{tab:DD Terms}
\end{table}
Once the images are corrected for dark, flat and smearing effects, the following \textit{differences} can be defined:
\begin{equation}
\Delta A=0.5 \cdot \mathrm{T_{A}}   \cdot  \{ I * \delta_\mathrm{I} + P * \delta_\mathrm{P} \} * S_\mathrm{A} + \Delta N_\mathrm{A},
\label{eq:delta_A}
\end{equation}
\begin{equation}
\Delta B=0.5 \cdot \mathrm{T_{B}}  \cdot  \{ I * \delta_\mathrm{I} - P * \delta_\mathrm{P} \} * S_\mathrm{B} + \Delta N_\mathrm{B},  
\label{eq:delta_B}
\end{equation}
where $\delta_\mathrm{I}$ and $\delta_\mathrm{P}$ refer to differences between the left and right beams. The difference between $\mathrm{T_{L}}$ 
and $\mathrm{T_{R}}$ produces a bias between left and right images, while the difference between $\mathrm {PSF_{L}}$ and $\mathrm{PSF_{R}}$ 
differentially distorts the images. In case the beamsplitter does not produce two identical beams (i.e. due to optical imperfections) 
$\delta_\mathrm{I}$ will be non-zero. Since ExPo looks for polarized intensities of $P \approx 10^{-5} I$, this term should be on the order of $\approx 10^{-4}$ 
or smaller, as will be explained in the next paragraphs. As soon as this condition is not satisfied, the polarization values will be affected by the uncorrected 
intensity differences.

\begin{figure}
  \centering
\includegraphics[width=\columnwidth,trim = 80 330 50 45]{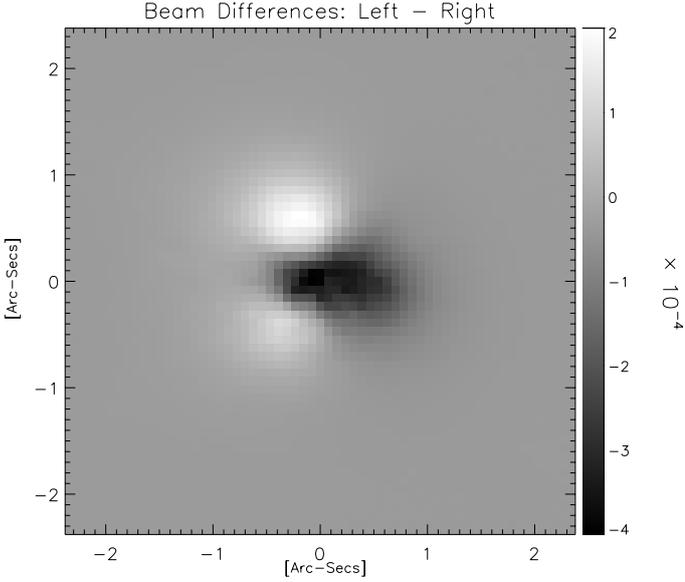}
   \caption{Normalized difference between left and right beams. The asymmetrical pattern around the center means \textit{shape} differences (i.e. $\mathrm{PSF_{L}}$ 
   and $\mathrm{PSF_{R}}$ produce different patterns). The color bar is in units of $10^{-4}$, meaning that this is the magnitude of the beam differences.}
  \label{fig:deltapsf}
\end{figure}

To evaluate the effect of $\delta_\mathrm{I}$ on the polarimetry results, two average intensity "left" and "right" images were produced by averaging left and right beams:
$I_\mathrm{L} = A_\mathrm{L} + B_\mathrm{L}, I_\mathrm{R} = A_\mathrm{R} + B_\mathrm{R}$. Both images were aligned with a third of a pixel accuracy, then normalized 
and finally subtracted. The noise contribution can be neglected in this calculation, since its contribution is several orders of magnitude below the intensity level.
The result is described by:
\begin{equation}
\centering
I_\mathrm{L} - I_\mathrm{R} =  (I*\delta^{'}_\mathrm{P} + P * \delta^{'}_\mathrm{I} ) * \delta_\mathrm{I},
\label{eq:intdiff}
\end{equation}
where $\delta^{'}_\mathrm{I}$ and $\delta^{'}_\mathrm{P}$ refer to any difference between the A and B frames. The 
difference between $\mathrm{T_{A}}$ and $\mathrm{T_{B}}$ will affect the whole image like a bias value, and the difference between $S_\mathrm{A}$ and $S_\mathrm{B}$ 
will affect the image shape. The later will be averaged out when combining enough images, i.e. the \textit{seeing} effect will affect the A
and B frames in the same way. FLC transmission differences are not higher than $10^{-3}$ for our current FLC \citep{Rodens_2011}, 
therefore $\delta^{'}_\mathrm{I}  \approx 10^{-3}$, and $\delta^{'}_\mathrm{P}$ will be very close to unity $\delta_\mathrm{P} \approx 1$. Taking all of this into account, 
the first term of Eq.~\ref{eq:intdiff} is much larger than the second term: $I*\delta^{'}_\mathrm{P} \gg P * \delta^{'}_\mathrm{I}$, so Eq.~\ref{eq:intdiff} becomes: 
\begin{equation}
\centering
I_\mathrm{L} - I_\mathrm{R}   \approx  I* \delta_\mathrm{I}.
\label{eq:intdiff2}
\end{equation}
Figure~\ref{fig:deltapsf} illustrates this result for the diskless star HD122815, observed with the FLC oriented at $0^{\circ}$. 
The image is scaled to its maximum, i.e. the beam differences have an 
amplitude of $\approx \pm 4 \cdot 10^{-4}$, which means that this is the magnitude of $\delta_{I}$.  Similar values are found when repeating 
this experiment for different FLC orientations.
As can be clearly seen from these results, beamsplitter imperfections are an important effect that needs to be taken into account when dealing with contrast ratios 
of $10^{4}$ or higher. Any differences in the PSF of each beam can modify the final result when aiming for such high contrast ratios.
This contribution is, however, minimized when working with the \textit{double-difference} method:
\begin{equation}
\sum_i^n (\Delta A - \Delta B)   =  \sum_{i}^{n/2} \left(I*\delta_\mathrm{I}*\delta^{'}_\mathrm{I}  + P*\delta_\mathrm{P}*\delta^{'}_\mathrm{P} + \Delta N_\mathrm{A} - \Delta N_\mathrm{B}\right).
\label{eq: doublediff}
\end{equation}
In this case, the four images produced in each FLC cycle are first combined, and then averaged over the whole data set. Now the intensity term is convolved 
with the convolution of $\delta_\mathrm{I}$ and $\delta^{'}_\mathrm{I}$. The latter will have a magnitude of $\delta_\mathrm{I} * \delta^{'}_\mathrm{I} \approx 10^{-7}$, 
so the intensity will be decreased by this factor. Both $\delta_\mathrm{P}$ and $\delta^{'}_\mathrm{P}$ are close to unity, so their convolution will be 
also close to unity. The polarization term is then not increased nor decreased. Finally, the noise term is not modulated by any 
\textit{delta} factor. Its contribution is minimized by the \textit{double-difference}, and it will be averaged out. This term will define the maximum contrast attainable with ExPo.

\begin{figure*}
  \centering
   	\resizebox{\hsize}{!}{\includegraphics[trim = 30 405 10 140]{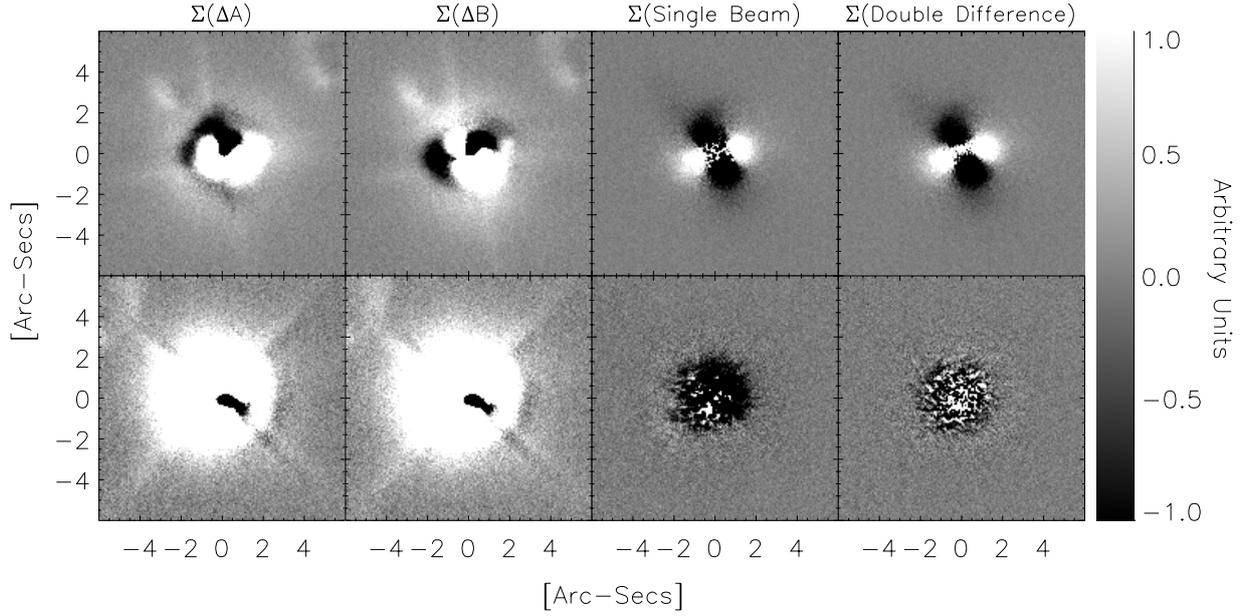}}
   \caption{Uncalibrated Polarized Image. Top rows: AB Aurigae. Bottom rows: Unpolarized Star. Color bar: (uncalibrated) polarization, P (Arbitrary Units). The results plotted 
   in the left and center columns are contaminated by instrumental artifacts. These artifacts are minimized when applying the \textit{double-difference}, as is shown in the right
    column.}
  \label{fig:double_diff}
\end{figure*}

Figure~\ref{fig:double_diff} summarizes the previous results for both a diskless star (HD122815) and a protoplanetary disk (AB Aurigae). The first two columns
starting from the left show the result given by Eq.~\ref{eq:delta_A} and Eq.~\ref{eq:delta_B}, respectively, when applied to these two targets. The result of a 
\textit{single-beam} experiment is shown in the third column. In this case, the images produce by just one beam (the right one) are shown to compare.
The result of the \textit{double-difference} described by Eq.~\ref{eq: doublediff} is shown in the right column. In the first and second columns the polarization pattern is contaminated 
by the intensity term described by $\delta_\mathrm{I}$, and the third column shows that the images are more noisy. The fourth column has no artifacts, showing a clean
polarization pattern, with the lowest noise level.
\begin{figure}
  \centering
   	\resizebox{\hsize}{!}{\includegraphics[trim = 90 430 75 145]{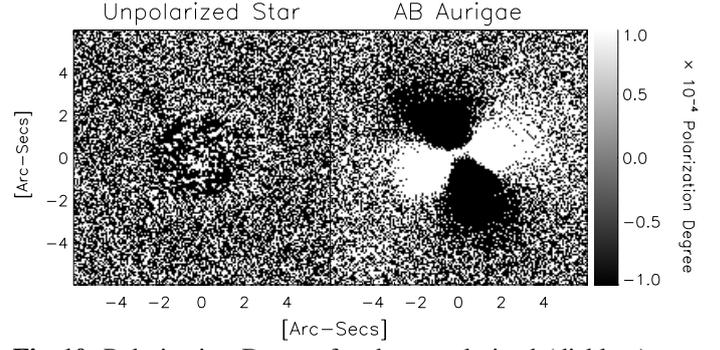}}
   \caption{Polarization Degree for the unpolarized (diskless) star HD122815 (left), and for AB Aurigae (right).}
  \label{fig:poldegree}
\end{figure}
\section{Additional Corrections}

\subsection{Instrumental Polarization}

The polarized light measured by ExPo contains not only the polarized light coming from the circumstellar environment, but also telescope polarization \citep{telescopepol} 
and interstellar polarization \citep{interstellarpol}. Moreover, any small difference between the beamsplitter transmission coefficients ($\mathrm{T_{A},T_{B}}$) or the FLC
transmission coefficients ($\mathrm{T_{L},T_{R}}$) introduces an artificial polarization signature. All these contributions are proportional to the intensity of the incoming light, 
so their effects can be measured by measuring the degree of polarization at the center of the (largely unpolarized) star. The polarization compensator used by ExPo minimizes
these effects, but there is always some leftover instrumental polarization. Taking into account that 
$A_\mathrm{L} + A_\mathrm{R} \approx \mathrm{T_{A}} \cdot S_\mathrm{A} * I * \delta_\mathrm{P}$, and 
$B_\mathrm{L} + B_\mathrm{R} \approx \mathrm{T_{B}} \cdot S_\mathrm{B} * I * \delta_\mathrm{P}$, the polarization degree 
at the star's center for the A frame can be described as:
\begin{equation}
\frac{A_\mathrm{L} - A_\mathrm{R}}{A_\mathrm{L} + A_\mathrm{R}}\Biggr|_{center}	 \approx	\frac{\delta_\mathrm{I}}{\delta_\mathrm{P}} +  \frac{P^{i}}{I} =  A_\mathrm{Bias},
\label{eq:A_Bias}
\end{equation}
and for the B frame:
\begin{equation}
\frac{B_\mathrm{L} - B_\mathrm{R}}{B_\mathrm{L} + B_\mathrm{R}}\Biggr|_{center} \approx	\frac{\delta_\mathrm{I}}{\delta_\mathrm{P}} -  \frac{P^{i}}{I} =  B_\mathrm{Bias}.
\label{eq: B_Bias}
\end{equation}
$P^{i}$ refers to any polarization feature that does not come from the circumstellar environment. As explained in \citet{Rodens_2011}, the polarization compensator is 
working for all observations (including unpolarized diskless stars), and it is impossible to set a true zero or reference level for our polarimetric observations. Therefore, 
all calculations are based on the assumption that the starlight is unpolarized. However this is not true in case there is strong inner disk polarization \citep{Canovas_2011_c}.
In these cases, a degree of polarization of the order of a few per cent is expected to be measured at the star's position. On the other hand, 
the expected instrumental polarization is a few percent as well, so in practice it is impossible to disentangle inner disk polarization from  instrumental polarization. 
The $A_\mathrm{Bias}$ and $B_\mathrm{Bias}$ are then subtracted from the \textit{degree of polarization} for the A and B frames:
\begin{equation}
\frac{P_\mathrm{A}}{I_\mathrm{A}} = \frac{A_\mathrm{L} - A_\mathrm{R}}{A_\mathrm{L} + A_\mathrm{R}} - A_\mathrm{Bias},
\label{eq: Eq12}
\end{equation}
\begin{equation}
\frac{P_\mathrm{B}}{I_\mathrm{B}} = \frac{B_\mathrm{L} - B_\mathrm{R}}{B_\mathrm{L} + B_\mathrm{R}} - B_\mathrm{Bias},
\label{eq: Eq13}
\end{equation}
where $dP_\mathrm{A}$ and $dP_\mathrm{B}$ are the instrumental polarization-corrected degree of polarization for the A and B frames, respectively. 
The (uncalibrated) polarization image $P_\mathrm{A}$ is  then calculated as the product of the degree of polarization times the intensity:
\begin{equation}
P_\mathrm{A} = \frac{P_\mathrm{A}}{I_\mathrm{A}} \cdot (A_\mathrm{L} + A_\mathrm{R}),
\label{eq: Eq14}
\end{equation}
\begin{equation}
P_\mathrm{B} = \frac{P_\mathrm{B}}{I_\mathrm{B}} \cdot (B_\mathrm{L} + B_\mathrm{R}).
\label{eq: Eq15}
\end{equation}
\begin{figure}
  \centering
	\includegraphics[width=\columnwidth,trim = 80 330 50 45]{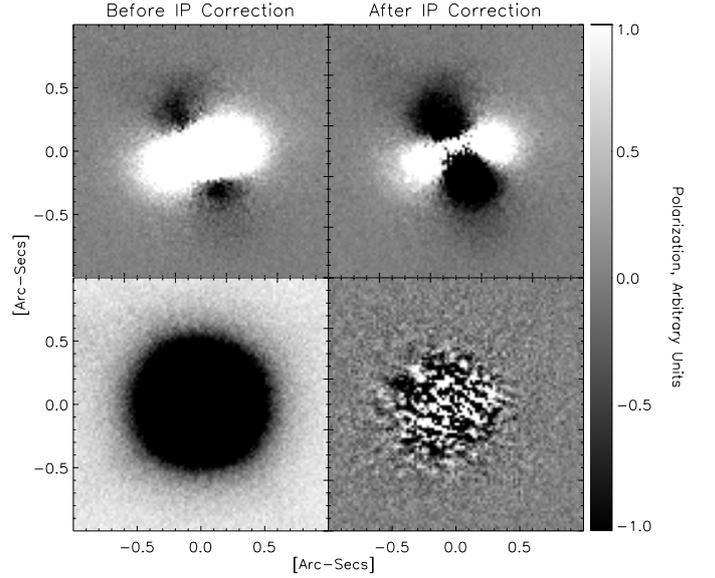}
   \caption{Effect of the instrumental polarization (IP) on the final images. Top: AB Aurigae. Bottom: Unpolarized star. Color bar: Polarization (P), arbitrary units.}
  \label{fig:instpol}
\end{figure}
Figure~\ref{fig:instpol} shows the effect of this correction.  An artificial bias appears in the polarization images when the IP correction is not applied. In case the star 
is unpolarized (bottom row), a bias level appears as a consequence of the imperfect compensation. In the case of AB Aur (top row) the polarization pattern is
distorted when no correction is applied.
\subsection{Field Rotation}
All alt-azimuth telescopes are affected by \textit{field rotation}. This term describes the amount of sky rotation, and it depends on the altitude ($h$) and azimuth ($a$) of the 
observed target, as well as the latitude ($\phi$) of the observer. It will reach its maximum value when pointing towards the zenith, and it will be almost negligible
when pointing to very low altitudes. If the observations are made at the Nasmyth focus, an extra rotation must be taken into account, due to the relative 
rotation between the telescope tube and the Nasmyth platform \citep[see][chapter4]{Marois_2006,Joos_Thesis}:
\begin{equation}
\dot{G} = \frac{ \cos{\phi} \cdot \cos{a} }{\cos{h}} \cdot \omega_{0},
\label{eq:frot}
\end{equation}
\begin{equation}
\dot{H} = - \sin{a} \cdot \cos{\phi} \cdot \omega_{0},
\label{eq:nasrot}
\end{equation}
\begin{equation}
\dot{F} = \dot{G}  \pm \dot{H},
\label{eq:totfieldrot}
\end{equation}
were $\omega_{0}$ is the \textit{sidereal rate}. $\dot{G}$ represents the rotation rate due to the sky movement, and $\dot{H}$
accounts for the rotation rate due to the telescope movement.
Equation~\ref{eq:totfieldrot} is the total amount of field rotation rate at the Nasmyth platform. The $\pm$ is related to the two platforms at the telescope. In the case of ExPo, 
the instrument has been placed at the "B" platform of the William Herschel Telescope, so the "+" applies.

While most of the instruments incorporate a de-rotator to remove this effect, the design of ExPo does not include one. It is necessary to calculate
$\dot{F}$ and then numerically rotate each observed target. Due to the short exposure time of the ExPo images, the amount of de-rotation 
between one image and the next image is too small to correct ($ \le 0.5^{\circ} $), so rotating each frame individually is not necessary. 
A rotation \textit{per image blocks} is performed instead. $\dot{F}$ is assumed to be constant within a certain interval. If the observed target is not close 
($\leq 15^{\circ}$) to the Zenith, this interval can be set to 30 seconds. Otherwise, shorter time intervals must be taken to minimize errors. 
Different rotation routines were tested. To test this,  different template images were first rotated and then de-rotated
by using different codes. The standard deviation $\sigma$ of the difference between the original template and the de-rotated version is a measurement
of the accuracy of the rotation procedure. This analysis shows that the best results are obtained when the image is first resampled to one fifth of a pixel. The rotation 
is then performed with the IDL routine \textit{rot}, using cubic interpolation, with the interpolation parameter set to -0.7.
\subsection{Sky Background Polarization} 
The polarization of the sky \citep{skypol} is variable, depending on the position of the target as well as the position of the Moon (since scattered moonlight is polarized). 
Unlike the instrumental polarization discussed in \S6.1, this contribution does not depend on the brightness of the observed target, but it behaves like a polarization
bias over the whole image. Therefore, the background polarization cannot be calculated by measuring the degree of polarization at the star's center.
The background polarization is subtracted at the very last stage of the data analysis, by measuring the polarization in the sky regions of the image. The median value of 
at least four different sky regions is calculated and subtracted from the reduced, polarization images.  
\subsection{Ghosts} 
\begin{figure}
  \center
	\resizebox{\hsize}{!}{\includegraphics[trim = 80 505 125 45]{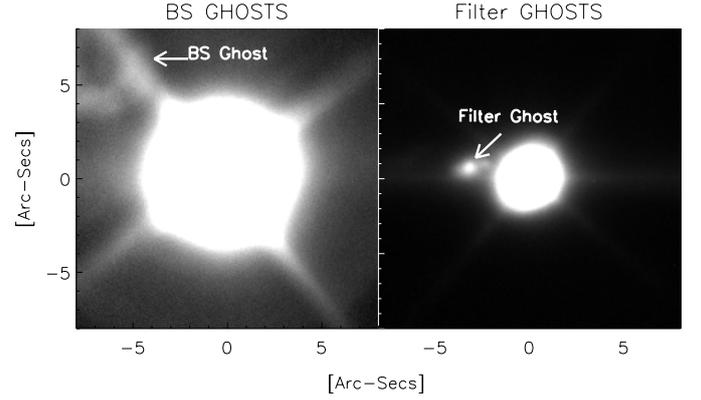}}
   \caption{Intensity images of the same star observed without (left) and with (right) $H_{\alpha cont}$ filter. The images are plotted at different contrast 
   in order to enhance the ghost image. The diffraction caused by the telescope spiders is clearly seen in the left image.}
  \label{fig: ghosts}
\end{figure}
Three different sources of optical ghosts are identified in ExPo:  the filters, the beamsplitter 
and the FLC. While the latter produces ghosts that move according to the FLC angle,  the  ghosts produced by filters and the beamsplitter
remain in a fixed position (an example of these ghosts can be seen in Fig.~\ref{fig: ghosts}). Filter ghosts are minimized by tilting the filters.
All this ghosts appears are contrast ratios of $10^{-4}$.

To analyze the effects caused by the ghosts on the polarized images, the mean and standard deviation ($\sigma$) was calculated around the position of the ghosts and 
around sky (empty) regions in the polarization images. This test was performed for several (ghost-affected) images, with the result that it is impossible 
to discriminate between ghosts and sky regions when looking at the polarized images, i.e., ghosts do not contribute to the polarization image any more than the sky
background. However, special care has to be taken when looking at the degree of polarization of the images. 
In this case, since the ghost appears in intensity (at contrast ratios of $\sim 10^{-4}$), it can contribute to the degree of polarization.
\section{Discussion}

Flat-field errors (i.e. pixel to pixel sensitivity variations) are not limiting ExPo. Since each image is shifted individually before averaging, 
flat-field effects are averaged out. However, flat-field effects might play an important role when an AO system is added to the instrument. In this case, the PSF
is expected to remain in a fixed position on the CCD, with no need to re-center the images.

Brightest-speckle alignment is not as good as template-alignment for our images. This result, in contradiction to previous work
\citep{Lucky_Thesis,Christou_1991} can be explained by the characteristics of this instrument. The small field of view used by ExPo, combined 
with the four different PSF's generated in each FLC cycle, makes a big difference with respect to the standard lucky-imaging observations. 
The speckle pattern produced by a single star is distributed over several pixels, and the beam differences make this pattern to be slightly
different in the left and right images, even though they correspond to simultaneous observations. As a result of this, the brightest speckle on the left
image is not necessarily the same as the brightest speckle on the right image. Therefore, the template center method produces better results than the speckle center
method. However, in case of good \textit{seeing} (below 0.7"), both methods converge, leading to very similar results. This can be explained by taking into
account that, in those cases, the PSF peak is concentrated in very few pixels, so the region covered by the speckle pattern is very small. Under these conditions,
the center defined by the template-center method and the brightest speckle are about the same.

While theoretically both the \textit{ratio} and the \textit{difference} method should produce similar results, this analysis shows that the first one fails 
when applied to a dual-beam imaging polarimeter like ExPo. The photon noise variance is signal dependent, i.e. it is proportional to the amount of photons 
in each image. On the other hand, each image produced by this instrument is affected by a different combination of beamsplitter and FLC transmission 
coefficients, which means that the total amount of light in each of the four images produced in one FLC cycle is slightly different.
This also applies to the photon noise, resulting in different levels of noise in each image. The \textit{double-difference} minimizes this effect, while
the \textit{dual-ratio} can increase it. This is a particular result for this experiment, but it can be generalized to any dual-beam polarimeter.

Numerical simulations show that the bias correction described in \S6.1 must be applied even when there is no instrumental polarization \citep{Canovas_2011_c}.
The strong polarization produced by the inner rim of a protoplanetary disk dominates the polarized intensity, making it much harder to detect the remaining circumstellar 
polarization.

As mentioned in \S2, future versions of ExPo may make use of 3 FLCs in a Pancharatnam configuration to minimize the chromatic effects 
introduced by the current, single-FLC configuration. This change will not affect the current data reduction process, but it will improve the instrument 
polarimetric performance.

The inclusion of an AO system in ExPo will lead to a much sharper PSF. The amount of speckles in these new PSFs will be considerably smaller than 
in our current PSFs. Real speckle-sized polarized features (i.e. exoplanets) might be removed as a result of the current CR rejection algorithm. One possible 
solution to this problem might be a new technique based on the comparison of the two \textit{differences} ($\Delta A$ and $\Delta B$) described by Eq.~\ref{eq:delta_A}
and Eq.~\ref{eq:delta_B}. A polarized speckle-sized  feature such as an exoplanet is expected to appear in both images at the same coordinates, while in case 
of a CR, it will appear in only one of these differences.

Chromatic effects introduced by the FLC and the BS are out of the scope of this paper, but must be taken into account in future instruments 
such as SPHERE or EPICS.

Finally, lucky imaging techniques (i.e. frame selection) does not produce significant improvements in our observations, since the standard exposure time for a 
target is around one hour (around 20 minutes per FLC position). In this regime,  the amount of ``lucky images" is not enough to produce a good signal to noise
ratio.

\section{Conclusion}

Dual-beam polarimeters are becoming more and more popular among the astronomical community, such as the ones described by \citet{Packham_2005}, 
\citet{Masiero_2008} and \citet{Nagaraju_2008}. The high contrast ratios achieved by polarimetry makes this technique a promising tool to characterize
exoplanets as well as circumstellar environments. 

ExPo can currently reach contrast ratios of $10^{5}$ at a four-meter class telescope, by means of polarimetry. As it is shown in this analysis, instrumental 
artifacts can be minimized by properly combining the data produced by a dual-beam instrument. The template centering method produces better results than the
speckle-centering method. The main limitations of the instrument are the beamsplitter imperfections and the FLC transmission differences, as explained in \S5. Future 
versions of this instrument will include an Adaptive Optics system as well as a coronagraph, which will require new approaches to the data reduction.

\begin{acknowledgements}
      We are grateful to all the staff at the William Herschel Telescope
      for their support when commissioning of ExPo. We also thank the referee for his useful comments and recommendations. 
\end{acknowledgements}

\bibliographystyle{aa.bst}	
\bibliography{biblio}		
%

%

\end{document}